\documentclass{ifacconf}

\usepackage{graphicx} 
\usepackage{natbib}        % required for bibliography
\usepackage{enumitem}
\usepackage{amsmath,amssymb,amsfonts}
\usepackage{xcolor}
\newtheorem{definition}[thm]{Definition}
\newtheorem{assumption}[thm]{Assumption}
\newtheorem{remark}[thm]{Remark}

\usepackage{tikz}
\usetikzlibrary{plotmarks}
\usetikzlibrary{patterns,ipe}

\usepackage{pgfplots}
\pgfplotsset{compat=newest,table/search path={./Figures/}}

\usepackage{environ}
\makeatletter
\newsavebox{\measure@tikzpicture}
\NewEnviron{scaletikzpicturetowidth}[1]{%
  \def\tikz@width{#1}%
  \begin{lrbox}{\measure@tikzpicture}%
  \BODY
  \end{lrbox}%
  \pgfmathparse{#1/\wd\measure@tikzpicture}%
  \BODY
}
\makeatother
 
\graphicspath{{./Figures/}}

\newcommand{\norm}[2]{\lVert#1\rVert_{#2}}
\newcommand{\abs}[1]{\left\lvert #1 \right\rvert}

\newcommand{\drawlinelegend}[1]{\raisebox{.5ex}{\tikz{\draw[#1, line width=0.4mm] (0,0) -- +(1em, 0);}}}
\usetikzlibrary{decorations.pathreplacing, patterns, calc, shapes,shapes.misc}
\tikzset{cross/.style={cross out, draw=black, minimum size=2*(#1-\pgflinewidth), inner sep=0pt, outer sep=0pt}, tcross/.default={1pt}}
\newcommand{\drawcross}[1]{\raisebox{.15ex}{\tikz{\draw[#1] (0,0) node[cross=2.5pt,#1] {};}}}
\newcommand{\drawcircle}[1]{\raisebox{.25ex}{\tikz{\draw[#1] (0,0) circle (0.5ex) ;}}}
\definecolor{mycolor1}{rgb}{0.00000,0.0,1.00000}%
\definecolor{mycolor2}{rgb}{0.00000,0.72202,0.00000}%
\definecolor{mycolor3}{rgb}{1.00000,0.0000,0.00000}%
\definecolor{mycolor4}{rgb}{0.00000,0.0000,0.00000}%
\begin{document}
\begin{frontmatter}

\title{Neural Network Training Using Closed-Loop Data: Hazards and an Instrumental Variable (IVNN) Solution\thanksref{footnoteinfo}} 
% Title, preferably not more than 10 words.

\thanks[footnoteinfo]{This work is supported by Topconsortia voor Kennis en Innovatie (TKI), and ASML and Philips Engineering Solutions.}

\author[First]{Johan Kon} 
\author[First,Second]{Marcel Heertjes}
\author[First,Third]{Tom Oomen} 
%\author[Third]{Third C. Author}

\address[First]{Control Systems Technology Group, Departement of Mechanical Engineering, Eindhoven University of Technology, P.O. Box 513, 5600 MB Eindhoven, The Netherlands, e-mail: j.j.kon@tue.nl}
\address[Second]{ASML, Mechatronics System Development, Veldhoven, The Netherlands}
\address[Third]{Delft Center for Systems and Control, Delft University of Technology, P.O. Box 5, 2600 AA Delft, The Netherlands}

\begin{abstract}                % Abstract of not more than 250 words.
%Measurement noise and disturbances may lead to inconsistent parameter estimates and limited performance when using closed-loop measurement data to train neural network feedforward controllers. The aim of this paper is to illustrate the occurrence of this inconsistency and to develop an estimation approach that is consistent. To address the closed-loop characteristics, a performance criterion based on instrumental variables is used. This criterion asymptotically recovers the true parameter values, given correct initialization. The hazards of training neural networks with closed-loop measurement data and the superiority of the proposed method are illustrated in a representative example.
An increasing trend in the use of neural networks in control systems is being observed. The aim of this paper is to reveal that the straightforward application of learning neural network feedforward controllers with closed-loop data may introduce parameter inconsistency that degrades control performance, and to provide a solution. The proposed method employs instrumental variables to ensure consistent parameter estimates. A nonlinear system example reveals that the developed instrumental variable neural network (IVNN) approach asymptotically recovers the optimal solution, while pre-existing approaches are shown to lead to inconsistent estimates.
%The results are considered important for many applications of neural networks in control systems.
\end{abstract}


\end{frontmatter}
%===============================================================================

\section{Introduction}
Improvements in feedforward control enable a major improvement in performance of control systems, e.g., in precision mechatronics. In feedforward control, two key requirements are typically considered \citep{Clayton2009}. First, high performance, i.e., a small tracking error, is desired. Second, task flexibility, i.e., performance for a variety of references, is required.

Learning techniques such as iterative learning control (ILC) \citep{Moore1993} have allowed the generation of feedforward signals that achieve performance that can compensate all reproducible behaviour \citep{Bristow2006,OomenAMC2020}. Despite this high performance, these approaches lack task flexibility, as is evidenced by the development of ILC for flexible tasks, such as a signal library  representing different subtasks \citep{Hoelzle2011}, and ILC with reference-dependent basis functions \citep{Wijdeven2010, Boeren2018}.

At the same time, traditional model-based feedforward control, which is highly flexible by design, has been further extended to enable higher performance. Advances include snap and static friction compensation through feedforward \citep{ChangACC2003, Boerlage2004, Kontaras2017, VanHaren2022} and rational feedforward to deal with higher-order flexibilities parametrizations \citep{6837472, Boeren2018}. These advances improve the performance and flexibility of the feedforward controller, but it is broadly experienced that these model-based extensions cannot achieve the same level of performance as learning-based techniques.

More flexible feedforward controller parametrizations have been investigated to go beyond the trade-off between performance and flexibility. For example, Gaussian processes are employed as non-parametric feedforward controllers in \cite{Blanken2020a, Poot2021}. Additionally, \cite{Aarnoudse2021, Bolderman2021, Kon2022PhysicsGuided} use neural networks as parametric feedforward controllers. In \cite{Bolderman2021, Kon2022PhysicsGuided}, neural networks are combined with a physics-based feedforward controller. In \cite{Aarnoudse2021}, these neural networks are trained using learned ILC input signals for a variety of references, and closed-loop control aspects are taken into account through a control-relevant cost function. A key advantage of closed-loop or ILC data is that nonlinearities manifest themselves along the trajectory of interest \citep{markusson2001, Schoukens2019}, thus enabling to train a feedforward controller that is accurate in the domain of interest. 
%and a regularization approach is used to ensure these are orthogonal to separate

Although neural networks have been shown to enable major performance improvements for feedforward control of closed-loop systems, at present the complete role of noise in closed-loop situations has not been investigated. The aim of this paper is to illustrate that this noise may lead to inconsistent parameter estimates when training neural networks based on closed-loop data, and to provide a solution resulting in consistent estimates.
% such as multilayer perceptrons and convolutional neural networks.% This is irrespective of whether data from ILC is used or regular closed-loop data. In fact, in the ILC case, the effect of noise is known to be amplified \cite[Section~IV.B.4]{OomenAMC2020}.

The results are illustrated for a specific class of systems and feedforward parametrizations for ease of exposition. The illustrated mechanism behind inconsistent estimates when using closed-loop data is independent of the specific neural network architecture, and holds for any neural network architecture, including convolutional neural networks and recurrent neural networks.

The main contribution of the present paper is an instru-mental-variables (IV) approach to train neural networks based on closed-loop data in the presence of input disturbances. This generalizes the parametric results in feedforward tuning \citep{Boeren2018} to the case where neural networks are used. This contribution consists of the following cornerstones. First, the neural network feedforward parametrization is introduced in Section \ref{sec:problem_formulation}. Second, in Section \ref{sec:least_squares_analysis}, it is shown that estimation of the parametrization's coefficients with a least-squares cost function on closed-loop data results in estimates that do not approach a minimizer that result in the best performance for increasing data. Third, an instrumental-variable cost function is introduced in Section \ref{sec:IV_criterion}, for which the estimates do (locally) converge to the true coefficients. Lastly, in Section \ref{sec:simulation_example}, these convergence properties and the consequences for performance are exemplified by simulation on a system with nonlinear friction characteristics.

\subsection*{Notation and Definitions}
For the finite-time signal $u$ with length $N$, $u(k) \in \mathbb{R}$ represents the signal at time index $k = 1,\ldots,N$, and $\underline{u} = \begin{bmatrix} u(1) & \ldots & u(N) \end{bmatrix}^T \in \mathbb{R}^N$ the vector representation of this finite time signal. 
%The expected value $\mathbb{E}(x) = \int_{-\infty}^{\infty} x f(x) dx$ represents the expectation of random variable $x$ with probability density function $f(x)$. 
% A coefficient estimate $\hat{\phi}(N)$ of $\phi_0$ on the basis of $N$ samples is said to be 
%unbiased if $\mathbb{E} \hat{\phi}(N) = \phi_0$, whereas it is said to be 
%consistent if it converges in probability to $\phi_0$, i.e., if $\lim_{N \rightarrow \infty} \textrm{Pr} \left( \abs{\hat{\phi}(N) - \phi_0} > \epsilon \right)  = 0 \ \forall \epsilon > 0$.

\section{Problem Formulation}
\label{sec:problem_formulation}
In this section, feedforward control for motion systems is introduced. Then, a neural network parametrization of the feedforward controller is given. The dataset to learn the coefficients of this parametrization is defined. Lastly, the problem of closed-loop noise in this dataset is formulated.
% Geen fan, misschien gewoon schrappen?
\begin{figure}
\centering
\includegraphics[width=\columnwidth]{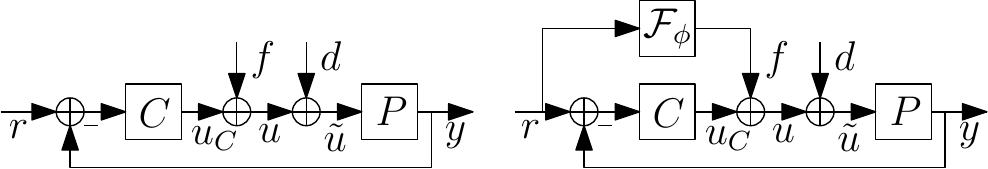}
\caption{Standard feedforward control setup with input additive noise $d$ (left) and feedforward control setup with neural network feedfoward controller $\mathcal{F}_\phi$ (right).}
\label{fig:FFW_setup}
\end{figure}

\subsection{Feedforward Control for Motion Systems}
The goal of feedforward control is to compensate for the effect of known exogenous inputs on the closed-loop system using an accurate feedforward signal. More specifically, consider the SISO control system shown in Fig. \ref{fig:FFW_setup} (left). The goal of feedforward control is to determine feedforward signal $f(k) \in \mathbb{R}$ such that the output $y(k) \in \mathbb{R}$ of the discrete-time plant $P$ equals the desired output $r(k) \in \mathbb{R}$ such that the error $e(k) \in \mathbb{R}$, given by
\begin{equation}
	e(k) = r(k) - y(k) = r(k) - P(f(k))
\end{equation}
is zero $\forall k$, with $k \in \mathbb{Z}_{\geq 0}$ the time index. The feedback controller $C$ aims to compensate the effects of both the unknown input disturbance $d$ and plant dynamics that are not compensated by the feedforward signal. 

If $P$ is linear time-invariant (LTI), the error is given by
\begin{equation}
	e = S r  + SP (f+d),
\end{equation}
in which $S = (1 + CP)^{-1}$ is the sensitivity function, such that
\begin{equation}
	f = P^{-1} r,
\end{equation}
achieves perfect tracking of the reference, i.e., $e=0$ in the noiseless ($d=0$) setting and under appropriate initial conditions. Since $P$ is not known exactly, various methods can be used to obtain $f$ such as manual tuning, ILC with basis functions \citep{6837472}, or system identification \citep{soderstromsysID} with inversion techniques \citep{VanZundert2018}.

However, $P$ usually contains nonlinear dynamics. More specifically, $P$ is an inverse nonlinear finite impulse response (NFIR) system, as is defined as follows.
\begin{definition}
	\label{def:plant}
	The plant $P$ with input $\tilde{u} \in \mathbb{R}$ satisfies the ordinary difference equation
	\begin{equation}
		P: \tilde{u}(k) \rightarrow y(k),\ \tilde{u}(k) = g_y(D_m(q^{-1})y(k)),
		\label{eq:plant_ODE}
	\end{equation}
	with $q^{-1}$ the forward shift operator, and
	\begin{equation}
		D_m(q)y(k) = \begin{bmatrix} y(k) & y(k-1) & \ldots & y(k-m) \end{bmatrix}^T,
	\end{equation}
	with $m \in \mathbb{Z}_{\geq 0}$ the maximum delay, and $g_y: \mathbb{R}^m \rightarrow \mathbb{R}$ an unknown, static, globally Lipschitz, nonlinear function.
\end{definition}
Existing LTI feedforward parametrizations cannot capture all relevant dynamics of $P$, resulting in a loss of performance and necessitating a broader parametrization.

%\begin{equation}
%P: \tilde{u}(k) = g_y(D(q) y(k))
%\end{equation}
%\begin{equation}
%D(q) = \begin{bmatrix} 1 & q^{-1} & q^{-2} & \ldots \end{bmatrix}^T
%\end{equation}

\subsection{Neural Network Feedforward Parametrization}
\begin{figure}
\centering
%\begin{scaletikzpicturetowidth}{2\columnwidth}
\definecolor{red}{rgb}{1,0,0}
\definecolor{blue}{rgb}{0,0,1}
\definecolor{green}{rgb}{0,1,0}
\definecolor{yellow}{rgb}{1,1,0}
\definecolor{orange}{rgb}{1,0.647,0}
\definecolor{gold}{rgb}{1,0.843,0}
\definecolor{purple}{rgb}{0.627,0.125,0.941}
\definecolor{gray}{rgb}{0.745,0.745,0.745}
\definecolor{brown}{rgb}{0.647,0.165,0.165}
\definecolor{navy}{rgb}{0,0,0.502}
\definecolor{pink}{rgb}{1,0.753,0.796}
\definecolor{seagreen}{rgb}{0.18,0.545,0.341}
\definecolor{turquoise}{rgb}{0.251,0.878,0.816}
\definecolor{violet}{rgb}{0.933,0.51,0.933}
\definecolor{darkblue}{rgb}{0,0,0.545}
\definecolor{darkcyan}{rgb}{0,0.545,0.545}
\definecolor{darkgray}{rgb}{0.663,0.663,0.663}
\definecolor{darkgreen}{rgb}{0,0.392,0}
\definecolor{darkmagenta}{rgb}{0.545,0,0.545}
\definecolor{darkorange}{rgb}{1,0.549,0}
\definecolor{darkred}{rgb}{0.545,0,0}
\definecolor{lightblue}{rgb}{0.678,0.847,0.902}
\definecolor{lightcyan}{rgb}{0.878,1,1}
\definecolor{lightgray}{rgb}{0.827,0.827,0.827}
\definecolor{lightgreen}{rgb}{0.565,0.933,0.565}
\definecolor{lightyellow}{rgb}{1,1,0.878}
\definecolor{black}{rgb}{0,0,0}
\definecolor{white}{rgb}{1,1,1}
\begin{tikzpicture}[ipe import]
  \node[ipe node]
     at (43.317, 758.342) {$r(k)$};
  \draw
    (76.022, 794.0117)
     -- (76.022, 782.0591)
     -- (96.1897, 781.9921)
     -- (96.1897, 793.9921)
     -- cycle;
  \node[ipe node]
     at (83.221, 785.157) {$1$};
  \draw
    (75.9271, 773.9959)
     -- (75.8323, 762.0433)
     -- (96.1897, 761.9921)
     -- (96.1897, 773.9921)
     -- cycle;
  \node[ipe node]
     at (78.462, 764.967) {$q^{-1}$};
  \draw[->]
    (48, 768)
     -- (76, 768);
  \draw[->]
    (68, 768)
     -- (68, 788)
     -- (76, 788);
  \draw[->]
    (68, 768)
     -- (68, 748)
     -- (76, 748);
  \draw[->]
    (96, 788)
     -- (112, 788);
  \draw[dashed]
    (64.1105, 797.956)
     -- (64, 732)
     -- (100, 732)
     -- (100.0943, 798.164)
     -- cycle;
  \node[ipe node, font=\footnotesize]
     at (64.719, 733.993) {$D_2(q^{-1})$};
  \draw[->]
    (96, 768)
     -- (112, 768);
  \draw[->]
    (96, 748)
     -- (112, 748);
  \draw[orange]
    (120, 788.208) circle[radius=8];
  \node[ipe node]
     at (114.879, 785.438) {$h_1^0$};
  \draw[seagreen]
    (120, 768) circle[radius=8];
  \node[ipe node]
     at (114.879, 765.438) {$h_2^0$};
  \draw[violet]
    (120, 748) circle[radius=8];
  \node[ipe node]
     at (114.879, 745.438) {$h_3^0$};
  \node[ipe node, font=\small]
     at (108, 808) {$\begin{gathered}
\textrm{Input} \\[-6pt]
\textrm{layer} \\[-6pt]
l = 0
\end{gathered}$};
  \draw
    (152, 788) circle[radius=8];
  \node[ipe node]
     at (146.879, 785.438) {$h_1^1$};
  \draw
    (152, 768) circle[radius=8];
  \node[ipe node]
     at (146.879, 765.438) {$h_2^1$};
  \draw
    (152, 748) circle[radius=8];
  \node[ipe node]
     at (146.879, 745.438) {$h_3^1$};
  \node[ipe node, font=\small]
     at (138.399, 808) {$\begin{gathered}
\textrm{Hidden} \\[-6pt]
\textrm{layer} \\[-6pt]
l =1
\end{gathered}$};
  \node[ipe node, font=\small]
     at (170.399, 808) {$\begin{gathered}
\textrm{Hidden} \\[-6pt]
\textrm{layer} \\[-6pt]
l = 2
\end{gathered}$};
  \draw[->]
    (160, 788)
     -- (176, 788);
  \draw[->]
    (160, 768)
     -- (176, 768);
  \draw[->]
    (160, 748)
     -- (176, 748);
  \draw
    (184, 788) circle[radius=8];
  \node[ipe node]
     at (178.879, 785.438) {$h_1^2$};
  \draw
    (184, 768) circle[radius=8];
  \node[ipe node]
     at (178.879, 765.438) {$h_2^2$};
  \draw
    (184, 748) circle[radius=8];
  \node[ipe node]
     at (178.879, 745.438) {$h_3^2$};
  \draw
    (216, 768) circle[radius=8];
  \draw[->]
    (192, 768)
     -- (208, 768);
  \node[ipe node]
     at (210.879, 765.438) {$h_1^3$};
  \draw[->]
    (224, 768)
     -- (240, 768);
  \node[ipe node]
     at (230.544, 773.712) {$f(k)$};
  \node[ipe node, font=\small]
     at (202.399, 808) {$\begin{gathered}
\textrm{Output} \\[-6pt]
\textrm{layer} \\[-6pt]
l = 3
\end{gathered}$};
  \draw[violet, ->]
    (128, 748)
     -- (144, 788);
  \draw[violet, ->]
    (128, 748)
     -- (144, 768);
  \draw[violet, ->]
    (128, 748)
     -- (144, 748);
  \draw[orange, ->]
    (128, 788)
     -- (144, 788);
  \draw[orange, ->]
    (128, 788)
     -- (144, 768);
  \draw[orange, ->]
    (128, 788)
     -- (144, 748);
  \draw[seagreen, ->]
    (128, 768)
     -- (144, 768);
  \draw[seagreen, ->]
    (128, 768)
     -- (144, 788);
  \draw[seagreen, ->]
    (128, 768)
     -- (144, 748);
  \draw[->]
    (160, 788)
     -- (176, 768);
  \draw[->]
    (160, 788)
     -- (176, 748);
  \draw[->]
    (160, 768)
     -- (176, 788);
  \draw[->]
    (160, 768)
     -- (176, 748);
  \draw[->]
    (160, 748)
     -- (176, 768);
  \draw[->]
    (160, 748)
     -- (176, 788);
  \draw[->]
    (192, 788)
     -- (208, 768);
  \draw[->]
    (192, 748)
     -- (208, 768);
  \draw[dashed]
    (102.3602, 798.0684)
     -- (102.3602, 731.9863)
     -- (228, 732)
     -- (228.0137, 798.164)
     -- cycle;
  \node[ipe node, font=\small]
     at (209.294, 734.052) {$NN$};
  \draw
    (75.9271, 753.996)
     -- (75.8323, 742.043)
     -- (96.1897, 741.992)
     -- (96.1897, 753.992)
     -- cycle;
  \node[ipe node]
     at (78.462, 744.967) {$q^{-2}$};
\end{tikzpicture}
%\end{scaletikzpicturetowidth}
\caption{Feedforward parametrization $\mathcal{F}_\phi$ as a combination of a two-lag delay line $D_2(q^{-1})$ and a multilayer perceptron with two fully connected layers.}
\label{fig:feedforward_parametrization}
\end{figure}
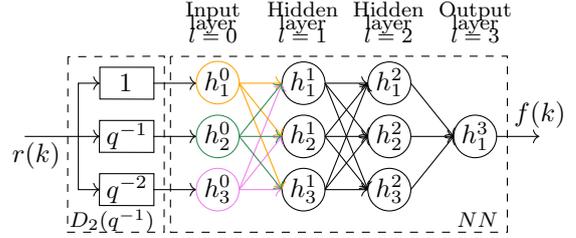

A neural network parametrization of the feedforward controller allows for learning all relevant plant dynamics from data, including unknown nonlinearities, since neural networks are universal approximators capable of approximating any nonlinear function. The feedforward controller $\mathcal{F}_\phi$ is parametrized by an NFIR system acting on the reference and its lags, see Fig. \ref{fig:FFW_setup} (right) and \ref{fig:feedforward_parametrization}, and is defined as follows.

\begin{definition}
The feedforward controller $\mathcal{F}_\phi$ satisfies
\begin{equation}	
\begin{aligned}
	\mathcal{F}_\phi: r(k) \rightarrow f(k),\ f(k) = \mathcal{F}_\phi(r(k)) = h^L(k),
	\label{eq:FFW_function}
\end{aligned}
\end{equation}
in which $h^L$ is the output of a neural network given by
\begin{align}
	h^l(k) &= D_p(q^{-1}) r(k) & \textrm{if}\ \ & l = 0 \nonumber \\
	h^l(k) &= \sigma \left( W^l z^{l-1}(k) + b^l \right) & \textrm{if}\ \  & l = {1, \ldots, L-1} \nonumber \\
	h^l(k) &= W^L z^{l-1}(k) + b^L & \textrm{if}\ \ & l = L, \label{eq:FNN}
\end{align}
with $W^l \in \mathbb{R}^{n_l \times n_{l-1}}$ the weights and $b^l \in \mathbb{R}^{n_l}$ the biases of layer $l$ with $n_l$ neurons, $\sigma(\cdotp)$ an element-wise activation function, such as a sigmoid or hyperbolic tangent, and full parameter set $\phi = \{W^l, b^l \}_{l=0}^{L}$.
\end{definition}
%\begin{bmatrix} r(k) & r(k-1) & \ldots & r(k-p) \end{bmatrix}^T
The feedforward class $\mathcal{F}_\phi$ in \eqref{eq:FFW_function} with $p = m$ can approximate $g_y$ in \eqref{eq:plant_ODE} up to arbitrary accuracy, such that $\abs{g_y(D_m(q^{-1}) y(k)) - \mathcal{F}_\phi(r(k))} < \epsilon \ \forall y(k) = r(k) \in \mathcal{R} \subset \mathbb{R}^m$ \citep{Goodfellow2016}. Given the correct parameters $\phi$, $\mathcal{F}_\phi$ can generate high performance feedforward signals for a variety of references.

% Remark: if delay present, just shift until relative degree is 0

\subsection{Closed-loop dataset}
Estimation of the feedforward parameters in \eqref{eq:FFW_function} requires a dataset describing the plant behaviour. A key advantage of closed-loop data over open-loop data is that the dataset captures the system dynamics around the trajectories of interest. Additionally, in many control applications, a feedback controller is required to operate the system due to open-loop instability or safety requirements. Therefore, a closed-loop dataset $\mathcal{D}$ covering the relevant IO space is considered, as formalized next.

\begin{definition}
	The dataset $\mathcal{D} = \{ \underline{u}, \underline{y} \},\ \underline{u}, \underline{y} \in \mathbb{R}^N$ contains finite-time closed-loop measurements satisfying
	\begin{align}
		y(k) = P(u(k)+d(k)) & \quad & u(k) = C(r(k)-y(k)).
	\end{align}
\end{definition}

A key disadvantage of a closed-loop dataset is that $u$ and $y$ contain spurious associations resulting from disturbance $d$, resulting in inconsistent estimates if not addressed appropriately.
\begin{remark}
Note that this closed-loop input data can also contain a contribution of a feedforward signal, obtained through, e.g., ILC. In fact, in the ILC case, the effect of disturbances is known to be amplified \citep[Section IV.B.4]{OomenAMC2020}, amplifying the inconsistency problem.
\end{remark}
\subsection{Problem Formulation}
The aim of this paper is to estimate $\phi$ of $\mathcal{F}_\phi$ in \eqref{eq:FFW_function} based on the closed-loop dataset $\mathcal{D}$ containing effects of input disturbance $d$ such that, in the disturbance-free case, $P\left(\mathcal{F}_\phi(r(k))\right) = r(k) \ \forall k \in \mathbb{Z}_{\geq 0}$. This includes
\begin{enumerate}[label=\arabic*)]
	\item illustrating that existing least-squares (LS) criteria result in inconsistent parameter estimates (Section \ref{sec:least_squares_analysis}).
	\item an IV criterion that results in consistent parameter estimates when initialized sufficiently close to the true parameters (Section \ref{sec:IV_criterion}), and
	\item a simulation example illustrating that inconsistent LS estimates deteriorate performance (Section \ref{sec:simulation_example}).
\end{enumerate}
% The aim of this paper is to illustrate that estimating $\phi$ based on a least-squares criterion produces inconsistent estimates due to confounding noise $d$ in a closed-loop dataset.
% The aim of this paper is develop an estimation criterion that produces consistent estimates $\phi$ based on closed-loop data to illustrate that estimates of $\phi$ based on a least-squares criterion produces inconsistent estimates due to confounding noise $d$ in a closed-loop dataset.

\section{Analysis of least-squares criterion with closed-loop data subject to noise}
\label{sec:least_squares_analysis}
This section illustrates that traditional approaches to neural network estimation can lead to poor estimation results. More specifically, it is shown that the inconsistency of the LS estimate also surfaces in the setting of neural network feedforward controllers estimated with closed-loop data. To this end, consider the LS criterion, which is commonly used for (nonlinear) regression in feedforward control with neural networks, defined as follows.
\begin{definition}
The least-squares criterion $J_{LS} \in \mathbb{R}_{\geq 0}$ is given by 
\begin{equation}
	J_{LS} = \sum_{k=1}^{N} \left( u(k) - \mathcal{F}_\phi(y(k)\right)^2 = \norm{\underline{u} - \mathcal{F}_\phi(\underline{y})}{2}^2,
	\label{eq:J_LS}
\end{equation}
in which $\mathcal{F}_\phi(\underline{y}) \in \mathbb{R}^N$ is shorthand notation for
\begin{equation}
	\mathcal{F}_\phi(\underline{y}) = \begin{bmatrix} \mathcal{F}_\phi(y(1)) & \mathcal{F}_\phi(y(2)) & \ldots & \mathcal{F}_\phi(y(N))	\end{bmatrix}^T.
\end{equation}
\end{definition}

% Wil ik dit al in S2?
For analysis purposes, it is assumed that feedforward parametrization \eqref{eq:FFW_function} contains the inverse plant, i.e., $P^{-1} \in \mathcal{F}_\phi$, as formalized by the following assumption.

\begin{assumption}
\label{ass:Pinv_in_model_set}
There exists a parameter set $\phi_0$ such that
\begin{equation}
	\mathcal{F}_{\phi_0}(P(u)) = u \ \forall u \in \mathbb{R}.
\end{equation}
\end{assumption}

This assumption is introduced merely to facilitate the forthcoming analysis. The main point of the analysis, i.e., the mechanism behind inconsistency for LS criterion \eqref{eq:J_LS}, is present irrespective of this assumption. In practice, this assumption can only be asymptotically satisfied for $u \in \mathcal{U} \subset \mathbb{R}$ and $N_\phi \rightarrow \infty$ by the universal approximation theorem \citep{Goodfellow2016}. Assumption \ref{ass:Pinv_in_model_set} enables the following lemma.
\begin{lem}
The least-squares criterion \eqref{eq:J_LS} is locally approximated around $\phi_0$ by $J^l_{LS} \in \mathbb{R}_{\geq 0}$ given by
\begin{equation}
	J^l_{LS} = \norm{\underline{u} - \mathcal{F}_{\phi_0}(\underline{y}) - F(\underline{y}) \Delta \phi}{2}^2,
	\label{eq:J_LS_local}
\end{equation}
with $F(\underline{y}) = \frac{\partial }{\partial \phi} \mathcal{F}_\phi(\underline{y}) \rvert_{\phi =\phi_0}$ and $\Delta \phi = \phi - \phi_0$.
\label{lem:local_LS_approximation}
\end{lem}

%\begin{pf}
%Consider a Taylor series expansion of $\mathcal{F}_\phi(\underline{y})$ around $\phi_0$, given by
%\begin{equation*}
%	\mathcal{F}_\phi(\underline{y}) = \mathcal{F}_{\phi_0}(\underline{y}) + \frac{\partial }{\partial \phi} \mathcal{F}_\phi(\underline{y}) \rvert_{\phi =\phi_0}(\phi - \phi_0) + \mathcal{O}\left( (\phi - \phi_0)^2 \right).
%\end{equation*}
%Neglecting second-order and higher terms, and substitution in \eqref{eq:J_LS} results in \eqref{eq:J_LS_local}, which completes the proof.
%\end{pf}

The following assumption ensures that \eqref{eq:J_LS_local} has a unique minimum.
\begin{assumption}
	$F^T(\underline{y}) F(\underline{y})$ is nonsingular.
	\label{ass:nonsingular_FtransposeF}
\end{assumption}
Assumption \ref{ass:nonsingular_FtransposeF} imposes a persistence of excitation condition on $y$, such that $\phi$ can be uniquely determined. Lemma \ref{lem:local_LS_approximation} and Assumption \ref{ass:nonsingular_FtransposeF} allow for expressing the estimate $\hat{\phi}_{LS}$.
\begin{thm}
\label{th:LS_local_optimum}
The estimate $\Delta \hat{\phi}_{LS} = \textrm{arg} \min_{\Delta \phi} J_{LS}^l$ is given by
\begin{equation}
\Delta \hat{\phi}_{LS} = \hat{\phi}_{LS} - \phi_0 = -\left( F^T(\underline{y}) F(\underline{y}) \right)^{-1} F^T(\underline{y}) \underline{d}.
\end{equation}
\end{thm}
%\begin{pf}
%$J_{LS}^l$ is quadratic in $\Delta \phi$, such that the solution is given by the pseudo-inverse, i.e.
%\begin{equation*}
%	\Delta \phi_{LS} = \left( F^T(\underline{y}) F(\underline{y}) \right)^{-1} F^T(\underline{y}) \left( \underline{u} - \mathcal{F}_{\phi_0}(\underline{y}) \right),
%\end{equation*}
%which is unique by Assumption \ref{ass:nonsingular_FtransposeF}. Furthermore, it holds that $\underline{y} = P(\underline{u} + \underline{d})$, such that
%\begin{equation*}
%\Delta \phi_{LS} = \left( F^T(\underline{y}) F(\underline{y}) \right)^{-1} F^T(\underline{y}) \left( \underline{u} - \mathcal{F}_{\phi_0}(P(\underline{u} + \underline{d})) \right),
%\end{equation*}
%which, by Assumption \ref{ass:Pinv_in_model_set}, reduces to
%\begin{equation*}
%	\Delta \phi_{LS} = \left( F^T(\underline{y}) F(\underline{y}) \right)^{-1} F^T(\underline{y}) \left( \underline{u} - \underline{u} - \underline{d} \right),
%\end{equation*}
%thereby completing the proof.
%\end{pf}
Next, the consistency of $\hat{\phi}_{LS}$ is investigated, in which consistency is defined as follows.
\begin{definition}
	A coefficient estimate $\hat{\phi}$ of $\phi_0$ based on $N$ samples is consistent if \citep{soderstromsysID}, with probability 1,
	\begin{equation}
		\lim_{N \rightarrow \infty} \hat{\phi} = \phi_0.
	\end{equation}
\end{definition}
A consistent estimate is both asymptotically unbiased and has diminishing variance for increasing data, i.e., converges to the true value with probability 1 for infinite data. Theorem \ref{th:LS_local_optimum} allows for analyzing the consistency of $\hat{\phi}_{LS}$, as formalized in the following theorem.
\begin{thm}
\label{cor:LS_is_biased_and_inconsistent}
%The expectancy of estimator $\Delta \hat{\phi}_{LS}(N)$ is given by
%\begin{equation}
%	\mathbb{E}\left(\Delta \hat{\phi}_{LS}(N)\right) = \mathbb{E} \left( -\left( F^T(\underline{y}) F(\underline{y}) \right)^{-1} F^T(\underline{y}) \underline{d} \right),
%\end{equation}
The least-squares estimate $\hat{\phi}_{LS}$ is inconsistent, i.e.,
\begin{equation}
	\lim_{N \rightarrow \infty} \hat{\phi}_{LS} - \phi_0 = \lim_{N \rightarrow \infty} \Delta \hat{\phi}_{LS} \neq 0.
\end{equation} 
\end{thm}
%\begin{pf}
%Under mild conditions \citep{soderstromsysID}, the limit of $\Delta \hat{\phi}_{LS}$ for infinite data is given by
%\begin{equation}
%\lim_{N \rightarrow \infty} \Delta \hat{\phi}_{LS} = R_{FF} ^{-1} R_{Fd},
%\end{equation}
%with $R_{FF}$ the autocorrelation of $F(y)$, and $R_{Fd}$ the cross correlation of $F(y)$ and $d$. The noise $d$ is also contained in $y$, such that $F(y)$ correlates with $d$, and thus, $R_{Fd} \neq 0$ and $\lim_{N \rightarrow \infty} \Delta \hat{\phi}_{LS} \neq 0$, completing the proof.
%\end{pf}
% \mathbb{E}\left(\Delta \hat{\phi}_{LS}(N)\right) \neq 0$

Theorem \ref{cor:LS_is_biased_and_inconsistent} shows that least-squares criterion \eqref{eq:J_LS} is inconsistent: even if the parametrization enables capturing the plant inverse, and $\hat{\phi}_{LS}$ is initialized sufficiently close to $\phi_0$ (or even at $\phi_0$), $\phi_0$ is not obtained as minimizer. Moreover, increasing noise levels increase the asymptotic bias in the parameter estimate.
 
%Nevertheless, the main point of the analysis, i.e., the mechanism behind parameter bias for least-squares criterion \eqref{eq:J_LS}, remains present without these assumptions;

\section{Instrumental variable Neural Network}
\label{sec:IV_criterion}
In this section, a new criterion for neural network training is presented that leads to consistent estimates by employing instrumental variables. It is shown that this IV criterion asymptotically recovers the true parameters if the optimization is initialized sufficiently close to the true parameters. To this end, consider the IV criterion defined as follows.
\begin{definition}
\label{def:J_IV}
The instrumental-variables criterion $J_{IV} \in \mathbb{R}_{\geq 0}$ is given by
\begin{equation}
	J_{IV} = \sum_{k=1}^{N} \norm{z(k)^T \left( u(k) - \mathcal{F}_\phi(y(k) \right)}{2}^2,
	\label{eq:J_IV}
\end{equation}
with $z(k) \in \mathbb{R}^{1\times N_\phi}$, or equivalently
\begin{equation}
	J_{IV} = \norm{Z^T \left( \underline{u} - \mathcal{F}_\phi(\underline{y}) \right) }{2}^2,
\end{equation}
with $Z \in \mathbb{R}^{N \times N_\phi}$ the matrix representation of $z(k)$.
\end{definition}
The IV criterion is well-known in system identification and has demonstrated to be successful to deal with closed-loop issues \citep{soderstromsysID} as well as in control \citep{Boeren2018}. The key novelty is its incorporation in neural network estimation, where LS criterion \eqref{eq:J_LS} is standard practice. A key difference is the fact that neural networks are nonlinear models. To this end, a local analysis is performed in the remainder. 

\begin{lem}\label{lem:local_IV_approximation}
The instrumental-variables criterion \eqref{eq:J_IV} can be locally approximated around $\phi_0$ by $J^l_{IV} \in \mathbb{R}_{\geq 0}$ given by
\begin{equation}
	J^l_{IV} = \norm{Z^T \left( \underline{u} - \mathcal{F}_{\phi_0}(\underline{y}) - F(\underline{y}) \Delta \phi \right)}{2}^2.
	\label{eq:J_IV_local}
\end{equation}
\end{lem}
%\begin{pf}
%Equivalent to Lemma \ref{lem:local_LS_approximation}.
%\end{pf}
The following assumption ensures that \eqref{eq:J_IV_local} has a unique minimum.
\begin{assumption}
	$Z^T F(\underline{y})$ is nonsingular.
	\label{ass:IV_correlated_condition}
\end{assumption}
Assumption \ref{ass:IV_correlated_condition} implies that $Z$ should be correlated with $F(\underline{y})$. 
%\textcolor{red}{
Since $F(\underline{y})$ represents the parameter Jacobian of $\mathcal{F}_\phi$, i.e., it is a nonlinear operation on $y$, this is usually satisfied in practice by picking $z$ such that it correlates with $y$, as in the linear IV case \citep{soderstromsysID}. Degenerate cases, such as the case where $\mathcal{F}(\underline{y})$ is singular, could, in theory, occur, and require further analysis, but are considered outside the scope of this paper.
%} 
Lemma \ref{lem:local_IV_approximation} and \ref{ass:IV_correlated_condition} allow for expressing the estimate $\hat{\phi}_{IV}$ of $J_{IV}^l$.

\begin{thm}
\label{th:IV_local_optimum}
The optimum $\Delta \hat{\phi}_{IV} = \textrm{arg} \min_{\Delta \phi} J_{IV}^l$ is given by
\begin{equation}
\Delta \hat{\phi}_{IV} = \hat{\phi}_{IV} - \phi_0 = -\left( Z^T(\underline{y}) F(\underline{y}) \right)^{-1} Z^T \underline{d}.
\end{equation}
\end{thm}
%\begin{pf}
%$J_{IV}^l$ is a linear set of equations in $\Delta \phi$, such that the solution is given by
%\begin{equation*}
%	\Delta \phi_{IV} = \left( Z^T(\underline{y}) F(\underline{y}) \right)^{-1} Z^T \left( \underline{u} - \mathcal{F}_{\phi_0}(\underline{y}) \right),
%\end{equation*}
%which is unique by Assumption \ref{ass:IV_correlated_condition}. Furthermore, it holds that $\underline{y} = P(\underline{u} + \underline{d})$, such that
%\begin{equation*}
%\Delta \phi_{IV} = \left( F^T(\underline{y}) Z(\underline{y}) \right)^{-1} Z^T \left( \underline{u} - \mathcal{F}_{\phi_0}(P(\underline{u} + \underline{d})) \right),
%\end{equation*}
%which, by Assumption \ref{ass:Pinv_in_model_set}, reduces to
%\begin{equation*}
%	\Delta \phi_{IV} = \left( Z^T(\underline{y}) F(\underline{y}) \right)^{-1} Z^T \left( \underline{u} - \underline{u} - \underline{d} \right),
%\end{equation*}
%thus completing the proof.
%\end{pf}
Theorem \ref{th:IV_local_optimum} allows for analyzing the consistency of $\hat{\phi}_{IV}$ under the following additional assumption.
\begin{assumption}
	Instruments $z$ are uncorrelated with $d$.
	\label{ass:IV_uncorrelated_condition}
\end{assumption}
%\textcolor{red}{
The freedom that exists in the construction of $Z$ readily allows for satisfying Assumption \ref{ass:IV_uncorrelated_condition}, e.g., through choosing $Z$ as the plant output for a different noise realization, as the reference, or as a plant output predicted through a model.
%} 
Under Assumption \ref{ass:IV_uncorrelated_condition}, consistency of $\hat{\phi}_{LS}$ can be concluded, as formalized next.
\begin{thm}
\label{cor:IV_consistent}
Given Assumptions \ref{ass:IV_correlated_condition} and \ref{ass:IV_uncorrelated_condition}, the estimate $\hat{\phi}_{IV}$ is consistent, i.e., with probability 1,
\begin{equation}
	\lim_{N \rightarrow \infty} \hat{\phi}_{IV} - \phi_0 = \lim_{N \rightarrow \infty} \Delta \hat{\phi}_{IV} = 0.
\end{equation} 
\end{thm}
%\begin{pf}
%The limit of $\Delta \hat{\phi}_{IV}$ for infinite data equals 
%\begin{equation}
%\lim_{N \rightarrow \infty} \Delta \hat{\phi}_{IV} = R_{ZF} ^{-1} R_{Zd},
%\end{equation}
%with $R_{ZF}$ the cross-correlation of $F(y)$ and $Z$, and $R_{Zd}$ the cross-correlation of $Z$ and $d$. If $Z$ is uncorrelated with $d$, $R_{Zd} = 0$ and $\lim_{N \rightarrow \infty} \Delta \hat{\phi}_{IV} = 0$, completing the proof.
%\end{pf}

Theorem \ref{cor:IV_consistent} shows that instrumental-variables criterion \eqref{eq:J_IV} provides consistent estimates: if $\hat{\phi}_{IV}$ is initialized sufficiently close to $\phi_0$, $\phi_0$ is obtained as minimizer for infinite data.

\section{Simulation example}
\label{sec:simulation_example}
In this section, the inconsistent parameter estimates of the LS cost function \eqref{eq:J_LS} are illustrated on an example dynamic system satisfying Definition \ref{def:plant}. In contrast, it is shown that the IV criterion \eqref{eq:J_IV} produces consistent estimates when initialized sufficiently close to $\phi_0$. Furthermore, the consequences of the inconsistent estimate on the performance are illustrated. 

\subsection{Example System}
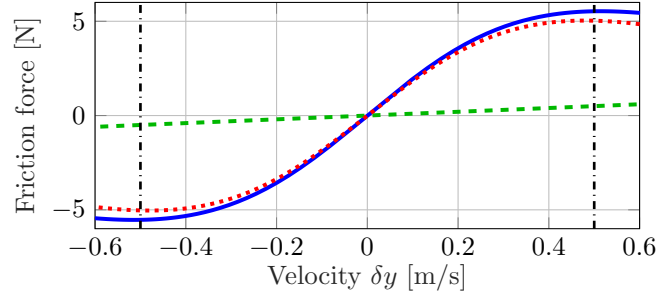
\begin{figure}
	\centering
	% This file was created by matlab2tikz.
%

%
\begin{tikzpicture}

\begin{axis}[%
width=7.1650cm,
height=3cm,
scale only axis,
xmin=-.6,
xmax=.6,
xlabel style={font=\color{white!15!black},yshift=.15cm},
xlabel={Velocity $\delta y$ [m/s]},
ymin=-6,
ymax=6,
ylabel style={font=\color{white!15!black}, yshift=-.1cm},
ylabel={Friction force [N]},
axis background/.style={fill=white},
xmajorgrids,
xminorgrids,
ymajorgrids,
yminorgrids
]
\addplot [color=mycolor1, line width=1.5pt, forget plot]
  table[]{SmoothStribeck-1.tsv};
\addplot [color=mycolor2, dashed, line width=1.5pt, forget plot]
  table[]{SmoothStribeck-2.tsv};
\addplot [color=mycolor3, dotted, line width=1.5pt, forget plot]
  table[]{SmoothStribeck-3.tsv};
\addplot [color=mycolor4, dashdotted, line width=1pt, forget plot]
  table[]{SmoothStribeck-4.tsv};
\addplot [color=mycolor4, dashdotted, line width=1pt, forget plot]
  table[]{SmoothStribeck-5.tsv};
\end{axis}
\end{tikzpicture}%
	\caption{Stribeck-like friction curve (\protect \drawlinelegend{mycolor1}) of example system \eqref{eq:smooth_stribeck_ODE} with $c_1 = 1$, $c_2 = 20$, $\alpha = 2.5$, consisting of a linear (\protect \drawlinelegend{mycolor2,dashed}) and nonlinear contribution (\protect \drawlinelegend{mycolor3,dotted}). This nonlinearity can be approximated up to arbitrary accuracy by $\mathcal{F}_\phi$ on the domain covered by $r$ (\protect \drawlinelegend{mycolor4,dash dot}).}
	\label{fig:smooth_stribeck}
\end{figure}

The plant $P$ is given by a mass-damper system with Stribeck-like friction characteristics, i.e., the kind of characteristics found in for example stage systems for lithographic inspection tools having a linear guidance with ball bearings, for which a simple model is given by
\begin{equation}
	\tilde{u}(k) = m \delta^2 y(k) + c_1 \delta y(k) + \frac{c_2 - c_1}{\cosh\left(\alpha \delta y(k) \right)} \delta y(k),
	\label{eq:smooth_stribeck_ODE}
\end{equation}
with parameters $c_1 = 1$, $c_2 = 20$, $\alpha = 2.5$, $m=5$, and in which $\delta y(k) = T_s^{-1} (y(k) - y(k-1))$ represents the discrete-time derivative, with $T_s = 1/1000$. Note that $\delta^2 y(k)$ is a linear combination of $y(k)$ and its $q=2$ delayed instances $y(k-1)$ and $y(k-2)$, such that \eqref{eq:smooth_stribeck_ODE} satisfies Definition \ref{def:plant}. The friction characteristics are visualized in Fig. \ref{fig:smooth_stribeck}.

The feedforward controller is parametrized as a neural network with 2 hidden layers with 10 neurons each and $\tanh$ activation functions, three input neurons and one linear output neuron, i.e.,
\begin{equation}
	\mathcal{F}_{\phi}(r(k)) = W_2 \sigma(W_1 \sigma(W_0 T D_2(q^{-1}) r(k) + b_0) + b_1) + b_2,
	\label{eq:FFW_parametrization_example}
\end{equation}
with $\phi = \{W_0, W_1, W_2, b_0, b_1, b_2 \}$, $W_0 \in \mathbb{R}^{10 \times 3}$, $W_1 \in \mathbb{R}^{10 \times 10}$, $W_2 \in \mathbb{R}^{10 \times 1}$, $b_0, b_1 \in \mathbb{R}^{10}$, $b_2 \in \mathbb{R}$. $T \in \mathbb{R}^{3 \times 3}$ is a fixed matrix that transforms $D_2(q^{-1})r(k)$ into a derivative basis to alleviate training. 

The system \eqref{eq:smooth_stribeck_ODE} is in feedback configuration, see Fig. \ref{fig:FFW_setup}, with controller $C(z)$ given by
\begin{equation}
	C(z) = \frac{123.38z - 122.76}{z^2 - 1.908z + 0.91}.
\end{equation}
Additionally, the system is subject to disturbance $d$ with 
\begin{equation}
	d = H(z) \nu =  \frac{0.8048z^2 -1.61z + 0.8048}{z^2 - 1.57z + 0.65} \nu,
\end{equation}
with $\nu$ white noise such that $\mathbb{E}\{\nu\} = 0$, $\mathbb{E}\{\nu^2\} = \sigma_\nu^2$.

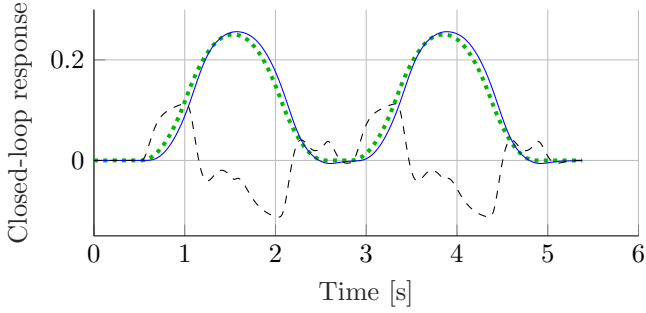
\begin{figure}
	\centering
	% This file was created by matlab2tikz.
%
\definecolor{mycolor1}{rgb}{0.00000,0.0,1.00000}%
\definecolor{mycolor2}{rgb}{0.00000,0.72202,0.00000}%
\definecolor{mycolor3}{rgb}{1.00000,0.0000,0.00000}%
\definecolor{mycolor4}{rgb}{0.00000,0.0000,0.00000}
\begin{tikzpicture}

\begin{axis}[%
width=7.1650cm,
height=3cm,
scale only axis,
xmin=0,
xmax=6,
xlabel style={font=\color{white!15!black}},
xlabel={Time [s]},
ymin=-0.15,
ymax=0.3,
ylabel style={font=\color{white!15!black}},
ylabel={Closed-loop response},
axis background/.style={fill=white},
axis x line*=bottom,
axis y line*=left,
xmajorgrids,
ymajorgrids
]
\addplot [color=mycolor2, forget plot, dotted,  line width=1.5pt]
  table[]{NominalOutput-1.tsv};
\addplot [color=mycolor1, forget plot]
  table[]{NominalOutput-2.tsv};
\addplot [color=black, forget plot, dashed]
  table[]{NominalOutput-3.tsv};
\end{axis}
\end{tikzpicture}%
	\caption{Nominal reference (\protect \drawlinelegend{mycolor2, dotted}), closed-loop output $y_0$ [m] (\protect \drawlinelegend{mycolor1}) and scaled input $u_0$ [N] (\protect \drawlinelegend{mycolor4, dashed}) for the noiseless setting.}
	\label{fig:reference_and_input}
\end{figure}

The reference to be tracked is a fourth-order reference with $v_{max} = 0.5$, $a_{max} = 1$, $j_{max} = 62$, $s_{max} = 4100$. Fig. \ref{fig:reference_and_input} shows the reference, the nominal noiseless output and the corresponding control effort (scaled).

To test the consistenty of $J_{LS}$ and $J_{IV}$, 20 noise realizations are generated for the standard deviation $\sigma_v$ ranging between $0$ to $0.01$. For each realization of the noise, a dataset consisting of the closed-loop responses $u,y$ to the above reference is created. Each dataset is used to estimate the parameters $\phi$ of the feedforward parametrization \eqref{eq:FFW_parametrization_example}, both according to LS criterion \eqref{eq:J_LS} and IV criterion \eqref{eq:J_IV}. For the IV criterion, the instrumental variables are given by the reference and its $N_\phi$ lags, i.e., 
\begin{equation}
	z(k) = \begin{bmatrix} r(k) & r(k-1) & \ldots & r(k-N_\phi+1)	\end{bmatrix}.
\end{equation} 
The parameters of each network are initialized at the parameters $\phi_0$ of a network trained until convergence in the noiseless setting, i.e., at a perfect inverse up to approximation capabilities.

\subsection{Parameter Inconsistency}
\begin{figure}[!t]
\centering
% This file was created by matlab2tikz.
%
\definecolor{mycolor1}{rgb}{0.00000,0.72202,0.00000}%
\begin{tikzpicture}

\begin{axis}[%
width=4.521in,
height=3.566in,
at={(0.758in,0.481in)},
scale only axis,
unbounded coords=jump,
xmin=0,
xmax=0.01,
xlabel style={font=\color{white!15!black}},
xlabel={Standard deviation of noise $\sigma_\nu$ [m]},
ymin=-0.3124,
ymax=-0.306,
ylabel style={font=\color{white!15!black}},
ylabel={Minimized coefficient $\phi_{3}$},
axis background/.style={fill=white},
axis x line*=bottom,
axis y line*=left,
xmajorgrids,
ymajorgrids,
width=6.56cm,
height=3.0cm,
scaled x ticks=false,
xticklabel style={/pgf/number format/fixed,/pgf/number format/precision=5},
scaled y ticks=false,
yticklabel=\pgfkeys{/pgf/number format/.cd,fixed,precision=3,zerofill}\pgfmathprintnumber{\tick},
xminorgrids,
yminorgrids
]
\addplot [color=black, line width=1.0pt, forget plot]
  table[]{Variation_coeff_3_LS_IV-1.tsv};
\addplot [color=red, only marks, mark=x, mark options={solid, red}, forget plot]
  table[]{Variation_coeff_3_LS_IV-2.tsv};
\addplot [color=red, only marks, mark=x, mark options={solid, red}, forget plot]
  table[]{Variation_coeff_3_LS_IV-3.tsv};
\addplot [color=red, only marks, mark=x, mark options={solid, red}, forget plot]
  table[]{Variation_coeff_3_LS_IV-4.tsv};
\addplot [color=red, only marks, mark=x, mark options={solid, red}, forget plot]
  table[]{Variation_coeff_3_LS_IV-5.tsv};
\addplot [color=red, only marks, mark=x, mark options={solid, red}, forget plot]
  table[]{Variation_coeff_3_LS_IV-6.tsv};
\addplot [color=red, only marks, mark=x, mark options={solid, red}, forget plot]
  table[]{Variation_coeff_3_LS_IV-7.tsv};
\addplot [color=red, only marks, mark=x, mark options={solid, red}, forget plot]
  table[]{Variation_coeff_3_LS_IV-8.tsv};
\addplot [color=red, only marks, mark=x, mark options={solid, red}, forget plot]
  table[]{Variation_coeff_3_LS_IV-9.tsv};
\addplot [color=red, only marks, mark=x, mark options={solid, red}, forget plot]
  table[]{Variation_coeff_3_LS_IV-10.tsv};
\addplot [color=red, only marks, mark=x, mark options={solid, red}, forget plot]
  table[]{Variation_coeff_3_LS_IV-11.tsv};
\addplot [color=red, only marks, mark=x, mark options={solid, red}, forget plot]
  table[]{Variation_coeff_3_LS_IV-12.tsv};
\addplot [color=red, only marks, mark=x, mark options={solid, red}, forget plot]
  table[]{Variation_coeff_3_LS_IV-13.tsv};
\addplot [color=red, only marks, mark=x, mark options={solid, red}, forget plot]
  table[]{Variation_coeff_3_LS_IV-14.tsv};
\addplot [color=red, only marks, mark=x, mark options={solid, red}, forget plot]
  table[]{Variation_coeff_3_LS_IV-15.tsv};
\addplot [color=red, only marks, mark=x, mark options={solid, red}, forget plot]
  table[]{Variation_coeff_3_LS_IV-16.tsv};
\addplot [color=red, only marks, mark=x, mark options={solid, red}, forget plot]
  table[]{Variation_coeff_3_LS_IV-17.tsv};
\addplot [color=red, only marks, mark=x, mark options={solid, red}, forget plot]
  table[]{Variation_coeff_3_LS_IV-18.tsv};
\addplot [color=red, only marks, mark=x, mark options={solid, red}, forget plot]
  table[]{Variation_coeff_3_LS_IV-19.tsv};
\addplot [color=red, only marks, mark=x, mark options={solid, red}, forget plot]
  table[]{Variation_coeff_3_LS_IV-20.tsv};
\addplot [color=red, only marks, mark=x, mark options={solid, red}, forget plot]
  table[]{Variation_coeff_3_LS_IV-21.tsv};
\addplot [color=red, line width=1.0pt, mark size=0.3pt, mark=*, mark options={solid, red}, forget plot]
  table[]{Variation_coeff_3_LS_IV-22.tsv};
\addplot [color=mycolor1, only marks, mark=o, mark options={solid, mycolor1}, forget plot]
  table[]{Variation_coeff_3_LS_IV-23.tsv};
\addplot [color=mycolor1, only marks, mark=o, mark options={solid, mycolor1}, forget plot]
  table[]{Variation_coeff_3_LS_IV-24.tsv};
\addplot [color=mycolor1, only marks, mark=o, mark options={solid, mycolor1}, forget plot]
  table[]{Variation_coeff_3_LS_IV-25.tsv};
\addplot [color=mycolor1, only marks, mark=o, mark options={solid, mycolor1}, forget plot]
  table[]{Variation_coeff_3_LS_IV-26.tsv};
\addplot [color=mycolor1, only marks, mark=o, mark options={solid, mycolor1}, forget plot]
  table[]{Variation_coeff_3_LS_IV-27.tsv};
\addplot [color=mycolor1, only marks, mark=o, mark options={solid, mycolor1}, forget plot]
  table[]{Variation_coeff_3_LS_IV-28.tsv};
\addplot [color=mycolor1, only marks, mark=o, mark options={solid, mycolor1}, forget plot]
  table[]{Variation_coeff_3_LS_IV-29.tsv};
\addplot [color=mycolor1, only marks, mark=o, mark options={solid, mycolor1}, forget plot]
  table[]{Variation_coeff_3_LS_IV-30.tsv};
\addplot [color=mycolor1, only marks, mark=o, mark options={solid, mycolor1}, forget plot]
  table[]{Variation_coeff_3_LS_IV-31.tsv};
\addplot [color=mycolor1, only marks, mark=o, mark options={solid, mycolor1}, forget plot]
  table[]{Variation_coeff_3_LS_IV-32.tsv};
\addplot [color=mycolor1, only marks, mark=o, mark options={solid, mycolor1}, forget plot]
  table[]{Variation_coeff_3_LS_IV-33.tsv};
\addplot [color=mycolor1, only marks, mark=o, mark options={solid, mycolor1}, forget plot]
  table[]{Variation_coeff_3_LS_IV-34.tsv};
\addplot [color=mycolor1, only marks, mark=o, mark options={solid, mycolor1}, forget plot]
  table[]{Variation_coeff_3_LS_IV-35.tsv};
\addplot [color=mycolor1, only marks, mark=o, mark options={solid, mycolor1}, forget plot]
  table[]{Variation_coeff_3_LS_IV-36.tsv};
\addplot [color=mycolor1, only marks, mark=o, mark options={solid, mycolor1}, forget plot]
  table[]{Variation_coeff_3_LS_IV-37.tsv};
\addplot [color=mycolor1, only marks, mark=o, mark options={solid, mycolor1}, forget plot]
  table[]{Variation_coeff_3_LS_IV-38.tsv};
\addplot [color=mycolor1, only marks, mark=o, mark options={solid, mycolor1}, forget plot]
  table[]{Variation_coeff_3_LS_IV-39.tsv};
\addplot [color=mycolor1, only marks, mark=o, mark options={solid, mycolor1}, forget plot]
  table[]{Variation_coeff_3_LS_IV-40.tsv};
\addplot [color=mycolor1, only marks, mark=o, mark options={solid, mycolor1}, forget plot]
  table[]{Variation_coeff_3_LS_IV-41.tsv};
\addplot [color=mycolor1, only marks, mark=o, mark options={solid, mycolor1}, forget plot]
  table[]{Variation_coeff_3_LS_IV-42.tsv};
\addplot [color=mycolor1, dashed, line width=1.0pt, mark size=0.3pt, mark=*, mark options={solid, mycolor1}, forget plot]
  table[]{Variation_coeff_3_LS_IV-43.tsv};
\end{axis}
\end{tikzpicture}%
\caption{Converged coefficient $(3,1)$ of $W_0$ for $\hat{\phi}_{LS}$ (\protect \drawcross{mycolor3}) and $\hat{\phi}_{IV}$ (\protect \drawcircle{mycolor2}) for 20 noise realizations for a range of standard deviations $\sigma_v$, with means (\protect \drawlinelegend{mycolor3}, \protect \drawlinelegend{mycolor2}) visualized for convenience. Here, $\hat{\phi}_{LS}$ does not correspond to the noiseless estimate (\protect \drawlinelegend{mycolor4, dashed}), i.e., is biased, and has large variance, and thus is inconsistent. In contrast, $\hat{\phi}_{IV}$  is consistent: $\hat{\phi}_{IV}$ corresponds to the noiseless estimate, i.e., does not contain bias, and has small variance.}
\label{fig:parameter_variation}
\end{figure}
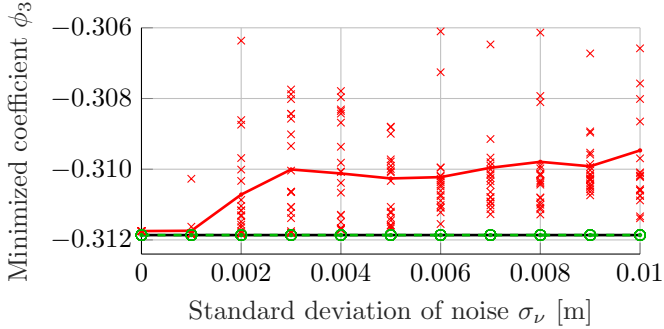

Based on above datasets, the (in)consistency of $\hat{\phi}_{LS}$ and $\hat{\phi}_{IV}$ is demonstrated. Fig. \ref{fig:parameter_variation} shows the converged parameter value of entry $(3,1)$ of $W_0$ for the criterion $J_{LS}$ and $J_{IV}$ for each noise realization over a range of the noise standard deviation $\sigma_\nu$.

From Fig. \ref{fig:parameter_variation}, the following is observed.
\begin{itemize}
	\item The estimate $\hat{\phi}_{LS}$ does not correspond to the true parameter $\phi_0$ of the noiseless setting for nonzero noise, since it is inconsistent. In contrast, $\hat{\phi}_{IV}$ is a consistent estimate and, thus, for small nonzero noise levels, does converge to $\phi_0$ up to minor (invisible) variations caused by the finite sample size $N$. 
	\item For $\sigma_\nu = 0$, both $\hat{\phi}_{LS}$ and $\hat{\phi}_{IV}$ converge to the same value .
	%up to solver precision, as there is no undermodelling of $P^{-1}$.
%	\item For noise levels $\sigma_\nu > 0.01$, the optimization converges to another local minimum corresponding to significantly different parameter values, such that the local consistency results no longer apply.
\end{itemize}

This inconsistency of $\hat{\phi}_{LS}$, i.e., deviation from the noiseless parameters, indicates that traditional criteria may lead to an inaccurate inverse in the presence of noise due to systematic errors, as will be shown next.
%This inconsistency of $\hat{\phi}_{LS}$, i.e., deviation from the noiseless parameters, illustrates that the learned feedforward filter does not capture the true inverse of $P$ for nonzero noise.

\subsection{Performance loss}
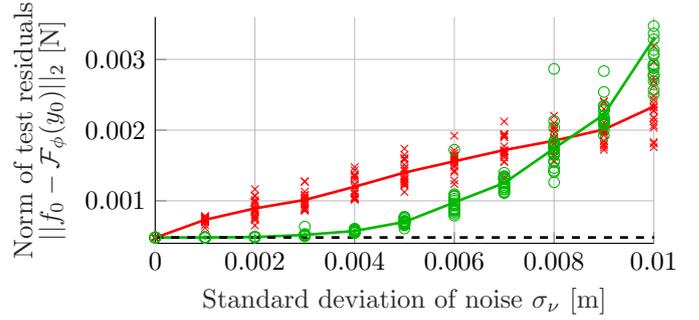
\begin{figure}[!t]
\centering
% This file was created by matlab2tikz.
%
\definecolor{mycolor1}{rgb}{0.00000,0.72202,0.00000}%
\begin{tikzpicture}

\begin{axis}[%
width=4.521in,
height=3.566in,
at={(0.758in,0.481in)},
scale only axis,
xmin=0,
xmax=0.01,
xlabel style={font=\color{white!15!black}},
xlabel={Standard deviation of noise $\sigma_\nu$ [m]},
ymin=0.0004,
ymax=0.0036,
ylabel style={font=\color{white!15!black}},
ylabel={Norm of test residuals $||f_0 - \mathcal{F}_{\phi}(y_0)||_2$ [N]},
axis background/.style={fill=white},
axis x line*=bottom,
axis y line*=left,
xmajorgrids,
ymajorgrids,
width=6.56cm,
height=3.0cm,
scaled x ticks=false,
xticklabel style={/pgf/number format/fixed,/pgf/number format/precision=5},
xlabel = {Standard deviation of noise $\sigma_\nu$ [m]},
scaled y ticks=false,
yticklabel={\pgfkeys{/pgf/number format/.cd,fixed,precision=3,zerofill}\pgfmathprintnumber{\tick}},
ylabel style={font=\color{white!15!black}, align=center},
ylabel={Norm of test residuals \\ $||f_0 - \mathcal{F}_{\phi}(y_0)||_2$ [N]},
xminorgrids,
yminorgrids
]
\addplot [color=red, only marks, mark=x, mark options={solid, red}, forget plot]
  table[]{Variation_test_cost_LS_IV_Initialization-1.tsv};
\addplot [color=red, only marks, mark=x, mark options={solid, red}, forget plot]
  table[]{Variation_test_cost_LS_IV_Initialization-2.tsv};
\addplot [color=red, only marks, mark=x, mark options={solid, red}, forget plot]
  table[]{Variation_test_cost_LS_IV_Initialization-3.tsv};
\addplot [color=red, only marks, mark=x, mark options={solid, red}, forget plot]
  table[]{Variation_test_cost_LS_IV_Initialization-4.tsv};
\addplot [color=red, only marks, mark=x, mark options={solid, red}, forget plot]
  table[]{Variation_test_cost_LS_IV_Initialization-5.tsv};
\addplot [color=red, only marks, mark=x, mark options={solid, red}, forget plot]
  table[]{Variation_test_cost_LS_IV_Initialization-6.tsv};
\addplot [color=red, only marks, mark=x, mark options={solid, red}, forget plot]
  table[]{Variation_test_cost_LS_IV_Initialization-7.tsv};
\addplot [color=red, only marks, mark=x, mark options={solid, red}, forget plot]
  table[]{Variation_test_cost_LS_IV_Initialization-8.tsv};
\addplot [color=red, only marks, mark=x, mark options={solid, red}, forget plot]
  table[]{Variation_test_cost_LS_IV_Initialization-9.tsv};
\addplot [color=red, only marks, mark=x, mark options={solid, red}, forget plot]
  table[]{Variation_test_cost_LS_IV_Initialization-10.tsv};
\addplot [color=red, only marks, mark=x, mark options={solid, red}, forget plot]
  table[]{Variation_test_cost_LS_IV_Initialization-11.tsv};
\addplot [color=red, only marks, mark=x, mark options={solid, red}, forget plot]
  table[]{Variation_test_cost_LS_IV_Initialization-12.tsv};
\addplot [color=red, only marks, mark=x, mark options={solid, red}, forget plot]
  table[]{Variation_test_cost_LS_IV_Initialization-13.tsv};
\addplot [color=red, only marks, mark=x, mark options={solid, red}, forget plot]
  table[]{Variation_test_cost_LS_IV_Initialization-14.tsv};
\addplot [color=red, only marks, mark=x, mark options={solid, red}, forget plot]
  table[]{Variation_test_cost_LS_IV_Initialization-15.tsv};
\addplot [color=red, only marks, mark=x, mark options={solid, red}, forget plot]
  table[]{Variation_test_cost_LS_IV_Initialization-16.tsv};
\addplot [color=red, only marks, mark=x, mark options={solid, red}, forget plot]
  table[]{Variation_test_cost_LS_IV_Initialization-17.tsv};
\addplot [color=red, only marks, mark=x, mark options={solid, red}, forget plot]
  table[]{Variation_test_cost_LS_IV_Initialization-18.tsv};
\addplot [color=red, only marks, mark=x, mark options={solid, red}, forget plot]
  table[]{Variation_test_cost_LS_IV_Initialization-19.tsv};
\addplot [color=red, only marks, mark=x, mark options={solid, red}, forget plot]
  table[]{Variation_test_cost_LS_IV_Initialization-20.tsv};
\addplot [color=red, line width=1.0pt, mark size=0.3pt, mark=*, mark options={solid, red}, forget plot]
  table[]{Variation_test_cost_LS_IV_Initialization-21.tsv};
\addplot [color=mycolor1, only marks, mark=o, mark options={solid, mycolor1}, forget plot]
  table[]{Variation_test_cost_LS_IV_Initialization-22.tsv};
\addplot [color=mycolor1, only marks, mark=o, mark options={solid, mycolor1}, forget plot]
  table[]{Variation_test_cost_LS_IV_Initialization-23.tsv};
\addplot [color=mycolor1, only marks, mark=o, mark options={solid, mycolor1}, forget plot]
  table[]{Variation_test_cost_LS_IV_Initialization-24.tsv};
\addplot [color=mycolor1, only marks, mark=o, mark options={solid, mycolor1}, forget plot]
  table[]{Variation_test_cost_LS_IV_Initialization-25.tsv};
\addplot [color=mycolor1, only marks, mark=o, mark options={solid, mycolor1}, forget plot]
  table[]{Variation_test_cost_LS_IV_Initialization-26.tsv};
\addplot [color=mycolor1, only marks, mark=o, mark options={solid, mycolor1}, forget plot]
  table[]{Variation_test_cost_LS_IV_Initialization-27.tsv};
\addplot [color=mycolor1, only marks, mark=o, mark options={solid, mycolor1}, forget plot]
  table[]{Variation_test_cost_LS_IV_Initialization-28.tsv};
\addplot [color=mycolor1, only marks, mark=o, mark options={solid, mycolor1}, forget plot]
  table[]{Variation_test_cost_LS_IV_Initialization-29.tsv};
\addplot [color=mycolor1, only marks, mark=o, mark options={solid, mycolor1}, forget plot]
  table[]{Variation_test_cost_LS_IV_Initialization-30.tsv};
\addplot [color=mycolor1, only marks, mark=o, mark options={solid, mycolor1}, forget plot]
  table[]{Variation_test_cost_LS_IV_Initialization-31.tsv};
\addplot [color=mycolor1, only marks, mark=o, mark options={solid, mycolor1}, forget plot]
  table[]{Variation_test_cost_LS_IV_Initialization-32.tsv};
\addplot [color=mycolor1, only marks, mark=o, mark options={solid, mycolor1}, forget plot]
  table[]{Variation_test_cost_LS_IV_Initialization-33.tsv};
\addplot [color=mycolor1, only marks, mark=o, mark options={solid, mycolor1}, forget plot]
  table[]{Variation_test_cost_LS_IV_Initialization-34.tsv};
\addplot [color=mycolor1, only marks, mark=o, mark options={solid, mycolor1}, forget plot]
  table[]{Variation_test_cost_LS_IV_Initialization-35.tsv};
\addplot [color=mycolor1, only marks, mark=o, mark options={solid, mycolor1}, forget plot]
  table[]{Variation_test_cost_LS_IV_Initialization-36.tsv};
\addplot [color=mycolor1, only marks, mark=o, mark options={solid, mycolor1}, forget plot]
  table[]{Variation_test_cost_LS_IV_Initialization-37.tsv};
\addplot [color=mycolor1, only marks, mark=o, mark options={solid, mycolor1}, forget plot]
  table[]{Variation_test_cost_LS_IV_Initialization-38.tsv};
\addplot [color=mycolor1, only marks, mark=o, mark options={solid, mycolor1}, forget plot]
  table[]{Variation_test_cost_LS_IV_Initialization-39.tsv};
\addplot [color=mycolor1, only marks, mark=o, mark options={solid, mycolor1}, forget plot]
  table[]{Variation_test_cost_LS_IV_Initialization-40.tsv};
\addplot [color=mycolor1, only marks, mark=o, mark options={solid, mycolor1}, forget plot]
  table[]{Variation_test_cost_LS_IV_Initialization-41.tsv};
\addplot [color=mycolor1, line width=1.0pt, mark size=0.3pt, mark=*, mark options={solid, mycolor1}, forget plot]
  table[]{Variation_test_cost_LS_IV_Initialization-42.tsv};
\addplot [color=black, dashed, line width=1.0pt, forget plot]
  table[]{Variation_test_cost_LS_IV_Initialization-43.tsv};
\end{axis}
\end{tikzpicture}%
\caption{Norm of residuals between true input $f_0$ and generated input $\mathcal{F}_\phi(y_0)$ for $\hat{\phi}_{LS}$ (\protect \drawcross{mycolor3}) and $\hat{\phi}_{IV}$ (\protect \drawcircle{mycolor2}) with noiseless output $y_0$ as test reference, for 20 noise realizations for a range of standard deviations $\sigma_v$. The inconsistency of $\hat{\phi}_{LS}$ causes a deviation of $\mathcal{F}_{\phi_{LS}}(y_0)$ from $f_0$. $\hat{\phi}_{IV}$ is locally consistent, and for small enough level, performs as well as the noiseless estimate 
(\protect \drawlinelegend{mycolor4, dashed}). }
\label{fig:performance_variation}
\end{figure}

The inconsistency of $\hat{\phi}_{LS}$ also results in a loss of performance. Fig. \ref{fig:performance_variation} shows the norm of residuals between the optimal input $f_0$ and the generated input $\mathcal{F}_\phi(y_0)$, in which $f_0$ and $y_0$ represent the plant input and output in the noiseless case to avoid extrapolation errors. 
%Fig. \ref{fig:performance_sample} shows these residuals for $\sigma_v = 0.005$ as a function of time for the median realization.

\begin{figure}[!t]
\centering
% This file was created by matlab2tikz.
%
\definecolor{mycolor1}{rgb}{0.00000,0.72202,0.00000}%
\begin{tikzpicture}

\begin{axis}[%
width=4.521in,
height=3.566in,
at={(0.758in,0.481in)},
scale only axis,
xmin=0,
xmax=5.378,
xlabel style={font=\color{white!15!black}},
xlabel={Time [s]},
ymin=-1,
ymax=1,
ylabel style={font=\color{white!15!black}},
ylabel={Force residual $f_0 - \mathcal{F}_\phi(y_0)$ [N]},
axis background/.style={fill=white},
axis x line*=bottom,
axis y line*=left,
xmajorgrids,
ymajorgrids,
width=6.56cm,
height=3.0cm,
scaled x ticks=false,
xticklabel style={/pgf/number format/fixed,/pgf/number format/precision=5},
scaled y ticks=true,
yticklabel=\pgfkeys{/pgf/number format/.cd,fixed,precision=2,zerofill}\pgfmathprintnumber{\tick},
ylabel style={font=\color{white!15!black}, align=center},
ylabel={Normalized residuals \\ $f_0 - \mathcal{F}_{\phi}(y_0)$ [-]},
xminorgrids,
yminorgrids
]
\addplot [color=red, forget plot]
  table[]{Median_f_LS_IV_0.005-1.tsv};
\addplot [color=mycolor1, forget plot]
  table[]{Median_f_LS_IV_0.005-2.tsv};
\end{axis}
\end{tikzpicture}%
\caption{Normalized residuals between true input $f_0$ and generated input $\mathcal{F}_{\hat{\phi}_{LS}}(y_0)$ (\protect \drawlinelegend{mycolor3}) and $\mathcal{F}_{\hat{\phi}_{IV}}(y_0)$ (\protect \drawlinelegend{mycolor2}), corresponding to the realization of median performance for noise level $\sigma_v = 0.005$. $\mathcal{F}_{\hat{\phi}_{LS}}(y_0)$ has larger deviations from the optimal input $f_0$ than $\mathcal{F}_{\hat{\phi}_{IV}}(y_0)$ due to the inconsistent estimates $\hat{\phi}_{LS}$.}
\label{fig:generated_input}
\end{figure}
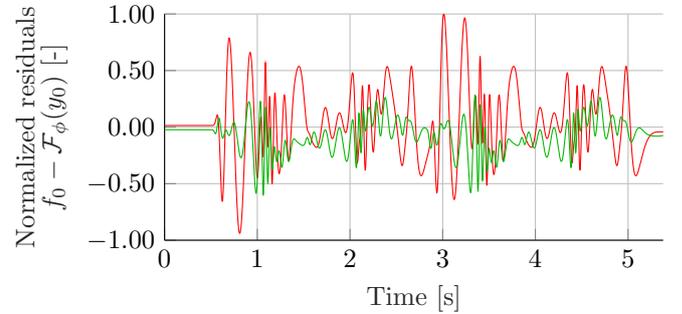

From Fig. \ref{fig:performance_variation}, the following is observed.
\begin{itemize}
	\item For zero noise, $\hat{\phi}_{LS}$ and $\hat{\phi}_{IV}$ result in the same performance. This is to be expected, since the estimates converge to almost exactly the same value (see Fig. \ref{fig:parameter_variation}) due to the low noise level. 
	\item For $\sigma_v \in (0, 0.008]$, $\hat{\phi}_{LS}$ suffers from reduced performance and more variation in performance due to the inconsistent estimates that depend on the noise realization. Additionally, higher noise levels result in more performance deterioration.
	\item For $\sigma_v > 0.002$, other effects besides (in)consistency of the estimator, discussed at the end of this section, start playing a role, such that the instrumental variable estimate is only locally able to obtain better estimates, and even performs worse for $\sigma_v > 0.008$. 
\end{itemize}

Additionally, Fig. \ref{fig:generated_input} shows the residuals between the optimal input signal $f_0$ and the generated feedforward signals $\mathcal{F}_{\hat{\phi}_{LS}}(y_0)$ and $\mathcal{F}_{\hat{\phi}_{IV}}(y_0)$ for the realization with median performance in terms of approximation norm of the optimal input for noise level $\sigma_\nu = 0.005$, see Fig. \ref{fig:performance_variation}. It is observed that $\mathcal{F}_{\hat{\phi}_{LS}}(y_0)$ is a worse approximation of $f_0$ than $\mathcal{F}_{\hat{\phi}_{IV}}(y_0)$ caused by inconsistency of $\hat{\phi}_{LS}$. These deviations from the optimal feedforward subsequently result in tracking errors.

Thus, the inconsistent estimate $\hat{\phi}_{LS}$ results in performance deterioration, and may be improved through instrumental variables. When the noise level invalidates these local results, additional phenomena start playing a role.
\begin{enumerate}[label=\arabic*)]
\item Extrapolation errors occur, since the input during training covers a different part of $g_y$ than the input during testing because the noise changes the trajectory of the closed-loop system. This requires that $\mathcal{F}_\phi$ extrapolates, and the quality of this extrapolation depends on the parameters. 
\item The quality of the local minima to which training converges, which depends on the noise realization, influences the performance. 
\item Most importantly, the choice of instruments represents an implicit weighting, potentially not penalizing residuals in specific time intervals, which reflects in the quality of the generated feedforward signal in these segments. More specifically, when the instruments are chosen as the reference and its delayed values, time segments of the residual for which this reference is 0 are not penalized, e.g., $t \in [0,0.5]$ s in Fig. \ref{fig:generated_input}. This allows for an offset in $\mathcal{F}_\phi(y_0)$ in these intervals (e.g., through the bias parameter of the last layer of $\mathcal{F}_\phi$), if only this offset results in smaller residuals for the penalized intervals. This is the mechanism behind the performance degradation of the IV criterion in Fig. \ref{fig:performance_variation}.
\end{enumerate}
These effects limit the applicability of the IV criterion outside a simulation context,  and require a global analysis, which is part of future research.

\section{Conclusion}
In this paper, it is shown that the straightforward application of the least-squares criterion for learning neural network feedforward controllers from closed-loop data may introduce inconsistent parameter estimates. Importantly, these errors may go unnoticed in naively training neural networks. Furthermore, it is shown that this inconsistency can result in performance deterioration in terms of the generated feedforward signal. An instrumental variable approach for training neural networks results in consistent parameter estimates, overcoming previous shortcomings, and results in superior feedforward performance when initialized sufficiently close to the true optimum. These results are exemplified by simulation on a representative system with nonlinear friction characteristics. Future work focuses on globally consistent parameter estimation for closed-loop data, addressing local minima, extrapolation errors, and the choice of instrumental variables.

\begin{ack}
The authors gratefully acknowledge the detailed discussions with Leontine Aarnoudse.
\end{ack}

\bibliography{../../Literature/Library.bib} 
                                                   
\end{document}